\documentclass[12pt]{iopart}

\usepackage{iopams}  

\usepackage{array}              
\usepackage{longtable}

\usepackage{hyperref}           
\usepackage{pdflscape}          
\usepackage{amsmath}            
\usepackage{rotating}           

\begin{document}

\title{L\'evy Light Cones and Critical Causality in Fractional Multiscale Quantum Ising Models}

\author{Joshua M Lewis$^{1,2}$, Zhexuan Gong$^{1,2}$ and Lincoln D Carr$^{1,2,3}$}

\address{$^1$Quantum Engineering Program, Colorado School of Mines, 1523 Illinois St., Golden, CO 80401, USA}
\address{$^2$Department of Physics, Colorado School of Mines, 1523 Illinois St., Golden, CO 80401, USA}
\address{$^3$Department of Applied Mathematics and Statistics, Colorado School of Mines, 1523 Illinois St., Golden, CO 80401, USA}

\ead{Joshualewis@mines.edu}

\begin{abstract}
We study causality and criticality in a one-dimensional fractional multiscale transverse-field Ising model, where fractional derivatives generate long range interactions beyond the scope of standard power laws. Such fractional responses are common in classical systems including the anomalous stress–strain behaviour of viscoelastic polymers, Lévy‑like contaminant transport in heterogeneous porous media, and the non‑Debye dielectric relaxation of glassy dielectrics. Furthermore, these unique interactions can be implemented in current quantum information architectures, with intriguing consequences for the many-body dynamics. Using a truncated Jordan-Wigner approach, we show that in the long wavelength limit of the mean field, the dynamical critical exponent is set by the fractional order q as $z=q/2$. To probe genuine many-body dynamics, we apply matrix-product-state simulations with the time-dependent variational principle adapted to nonlocal couplings.  Tracking the entanglement-entropy light cone and performing finite-size scaling of the many-body gap for $0<q<2.5$, we confirm a continuously tunable exponent $z(q)$: for $q<2$ the entanglement front broadens with a sublinear light cone; for $2<q<2.5$ we observe a faint superlinear cone indicative of $z<1$; and for $q \gtrsim 2.5$ the system reverts to the ballistic nearest-neighbour regime with $z=1$.  The correspondence between quantum entanglement fronts that spread as $t^{1/z}$ and classical Lévy flights whose mean-square displacement grows as $t^{2/q}$ provides a direct physical link between fractional interactions and Lévy statistics.  Fractional derivatives therefore offer a unified framework in which short-range, power-law, and frustrated long-range interactions emerge as limiting cases, enabling controlled exploration of nonlocal causality bounds and exotic entanglement dynamics within current quantum information platforms.
\end{abstract}



\maketitle

\section{Introduction}

Understanding how quantum information propagates in many-body systems is a central theme in quantum information science and technology (QIST), motivated by both foundational questions on non-equilibrium physics and practical considerations for quantum simulators/emulators. In systems with short-range interactions, Lieb and Robinson established rigorous bounds on the speed of information transfer, enforcing a linear light cone for correlation growth~\cite{lieb1972finite}. However, numerous quantum platforms---including trapped ions, Rydberg atoms, and ultracold atoms in optical lattices---exhibit long-range couplings that deviate substantially from these short-range models~\cite{jurcevic2014quasiparticle, zhang2017optical, richerme2014non}. Such power-law decays of the form $1/r^\alpha$ can give rise to faster-than-linear light cones and other exotic dynamical phenomena~\cite{hauke2013spread, gong2017entanglement, foss2015nearly, yang2021deconfined, yang2022topological, yang2022fate}.

Despite this progress in power-law decaying systems, many open questions remain regarding how genuinely nonlocal interactions reshape quantum information transfer, especially when the range and sign of the couplings vary beyond standard power-law forms. Examples of more intricate interaction profiles include Lennard-Jones potentials in chemistry, Yukawa interactions for screened fields, DLVO potentials for colloids, and various effective couplings in active matter. In these scenarios, the interplay of heavy-tailed distributions and potentially sign-alternating couplings can lead to dynamics far richer than those seen in purely ferromagnetic or anti-ferromagnetic power-law interacting spin models.

Fractional quantum mechanics provides a framework for tackling such nonlocal effects by introducing Riesz fractional derivatives $\partial_x^q \equiv \frac{\partial^q}{\partial x^q}$ into the Hamiltonian~\cite{laskin2000fractional, laskin2000fractionalQ, laskin2002fractional, stickler2013potential}. In momentum space, these derivatives act as Fourier multipliers $-|k|^q$, thereby interpolating continuously between short-range and long-range regimes. Crucially, this formulation emerges naturally from stable Lévy flights, a generalization of Brownian motion whose jump-length distributions remain heavy-tailed under convolution. Unlike artificially constructed $1/r^\alpha$ models, fractional derivatives encode both long-range and sign-alternating effects within a single operator. 

In single-particle quantum mechanics, extending the usual dispersive term to include fractional derivatives was pioneered by Laskin, who reinterpreted the Feynman path integral so that it encompasses L\'evy flights instead of standard Brownian paths~\cite{laskin2000fractional, laskin2000fractionalQ, laskin2002fractional, stickler2013potential, hasan2018tunneling, zhang2015propagation}. In the conventional Gaussian picture, typical paths exhibit Hausdorff (fractal) dimension $H_D = 2$ in a two-dimensional embedding, where $H_D$ is defined by the filling of a space when considering an infinite random walk. By contrast, L\'evy flights are governed by a stable distribution whose jump-length probability density falls off as $\mathcal{P}_q(x)\propto 1/|x|^{q+1}$, allowing $H_D$ to vary with the L\'evy stability index $q$. Because L\'evy distributions remain stable under convolution, these microscopic nonlocal jumps can produce robust, emergent macroscopic laws. Such phenomena have been studied in a variety of contexts, including dynamical correlations in low-dimensional Hamiltonian systems~\cite{mendl2015current, van2012exact, dhar2013exact, kundu2019fractional}, turbulence~\cite{solomon1993observation}, non-Newtonian fluids~\cite{pandey2016linking}, animal foraging patterns~\cite{brockmann2006scaling, benhamou2007many, murakami2019levy}, neuron signaling~\cite{liu2021levy}, and financial market data~\cite{yarahmadi20222d}. Experimental realizations of L\'evy statistics in designer materials built from the ground up have also emerged in optical platforms~\cite{iomin2021fractional, zeng2019one, xin2021propagation, he2021propagation, zhang2015propagation}, with recent demonstrations in L\'evy-based waveguide experiments~\cite{liu2023experimental}.

When translated to many-body settings, fractional derivatives can introduce a discrete spin--spin coupling $J(r)$ whose decay and sign depend on the fractional order $q$. These interactions are well characterized by a asymptotic power law, $r^{-(1+q)}$, with additional subleading power laws. Notably, for $q\le2$, the fractional multiscale model observes a sublinear light cone---while continuously matching the short-range limit at $q=2$. However, \emph{once $q>2$, the long-range couplings no longer behave as a mere decaying tail but instead manifest sign-alternating (frustrating) interactions.} This yields a subtle superlinear light cone and a departure from standard Lieb-Robinson-like bounds, pointing to a new regime of dynamics. Indeed, as we recently showed, fractional multiscale media can give rise to a new tunable universality class in both classical and quantum systems~\cite{lewis2025classical}.

Viewed from a Lévy‑flight perspective, the fractional multiscale transverse‑field Ising model is more than a convenient interpolation, it is the quantum analog of a classical Lévy walk.  Varying the fractional order $q$ reshapes the heavy‑tailed statistics encoded in the spin–spin coupling, continuously connecting the short‑range limit at $q=2$ to a genuinely multiscale regime.  Because those statistics stem from the same stable‑distribution framework that underlies anomalous transport in turbulence, finance, biology, and soft matter, the fractional multiscale transverse‑field Ising model offers a physically motivated backbone for long‑range quantum matter. This unified viewpoint links current experiments on trapped ions, Rydberg arrays, and cavity‑mediated cold atoms to the rich physical structure of Lévy processes, supplying clear, tunable predictions for how non‑Gaussian propagation and frustration should emerge in near‑term quantum simulators.

In this work, we investigate these multiscale phenomena through a one-dimensional fractional multiscale transverse-field Ising model, employing matrix product state (MPS) methods specifically adapted for long-range interactions~\cite{wall2009emergent,frowis2010tensor}. By utilizing the time-dependent variational principle (TDVP), we simulate the real-time evolution of localized perturbations and systematically analyze how fractional derivatives reshape causality. In particular, we uncover a superlinear light cone for $q>2$ and extract the corresponding dynamical critical exponent $z$, establishing direct links between fractional order and anomalous correlation spreading. Our results unify short-range, long-range, and frustrated regimes under a single framework, revealing how heavy-tailed Lévy statistics can control novel quantum dynamical phases. We conclude by discussing the implications of these findings for near-term quantum simulators and future directions in nonlocal quantum many-body physics.

\section{The Fractional Ising Model}

Historically, many classical and quantum field theories can be derived from the stochastic motion of virtual particles mediating interactions in a many-body system. In cases dominated by Brownian motion, the resulting interaction kernels are effectively local or rapidly decaying, allowing them to be approximated by second-derivative terms. However, when one extends to L\'evy flights with heavy-tailed distributions, long-range correlations naturally emerge. This distinction leads to fractional or otherwise nonlocal derivatives appearing in the effective Hamiltonian. 

A standard way to define the fractional derivative of order $q$ is through its action in momentum space. Specifically,
\begin{equation}
    \label{eq:FourierFracDeriv}
    \partial_x^q f(x) =\mathcal{F}^{-1}
    [
        -|k|^q\,\mathcal{F}(f(x))
    ] \,,
\end{equation}
where $\mathcal{F}$ is the Fourier transform. This definition makes explicit that $\partial_x^q$ is a nonlocal operator in position space, but corresponds to multiplying by $-|k|^q$ in momentum space.

L\'evy flights provide a natural origin for such terms. Their stable probability distribution $\mathcal{P}_q(x)\propto 1/|x|^{q+1}$ has the characteristic function $e^{-|k|^q}$, indicating that an infinite product of small-step propagators will accumulate a factor of $-|k|^q$ in the exponent independent of initial condition. As shown in fractional quantum mechanics~\cite{laskin2000fractional, laskin2000fractionalQ, laskin2002fractional}, this construction leads to an interaction kernel that departs from ordinary local behavior and instead captures multi-scale correlations reflective of the heavy-tailed distribution. When viewed through the lens of field theory, these L\'evy-based processes produce fractional derivative operators $\partial_x^q$ in the Hamiltonian. The construction of such materials can be engineered with artificially induced scattering mechanisms such as in optical systems \cite{barthelemy2008levy}.

Upon discretizing the fractional derivative induced by these L\'evy flights applied onto a many body system, one obtains a quantum spin chain in which the fractional derivative induces long-range interactions among lattice sites. Concretely, let $\hat{\sigma}_j^z$ and $\hat{\sigma}_j^x$ denote the usual Pauli operators at site $j$, and define $r=|i-j|$. We then write the fractional multiscale transverse-field Ising Hamiltonian as
\begin{equation}
    \hat{H} = -J_0 \sum_{i<j} J(r)\hat{\sigma}_i^x \hat{\sigma}_j^x + g \sum_{j} \hat{\sigma}_j^z \,,
    \label{eq:FracIsing}
\end{equation}
where $J_0$ sets the overall interaction scale and $g$ is the transverse-field strength; we treat the analog to the ferromagnetic case, $J_0>0$. The coupling $J(r)$ encodes nonlocal interactions derived from the Riesz fractional derivative, which can be implemented via a second-order discretization scheme \cite{ortigueira2006riesz}. 

For $q=2$, $J(r)$ reproduces the familiar nearest-neighbor or short-range limit, whereas $q>2$ yields sign-alternating, faster-decaying couplings that induce frustration. Conversely, $q<2$ corresponds to more strongly nonlocal interactions. The coupling function itself is given by
\begin{equation}
    J(r) = (-1)^{r + 1} \binom{q}{\frac{q}{2} + r} \,.
    \label{eq:coupling}
\end{equation}
However, for $q=2$ these asymptotic tails vanish. In this discrete setting, the resulting spin model captures how L\'evy-type nonlocalities reshape the competition between ordering ($\hat{\sigma}^x_i \hat{\sigma}^x_j$) and quantum fluctuations ($\hat{\sigma}^z_j$), thus generalizing the transverse-field Ising model to a genuinely fractional regime. The discrete coupling $J(r)$ behaves asymptotically like a power law with subleading power law corrections. Asymptotically, for large r the coupling behaves as
\begin{equation}
    J(r) \sim r^{-(1+q)} + r^{-(3+q)} \,,
    \label{eq:assymtoticCoupling}
\end{equation}
while short- to medium-range distances exhibit additional nontrivial local structure.

\section{Numerically Extracting the Dynamical Critical Exponent}

One-dimensional matrix product states (MPS) have become a powerful tool for studying quantum many-body systems. Their chief advantage lies in the ability to compactly represent states obeying area-law entanglement, such as the ground states of most gapped, one-dimensional Hamiltonians. As a result, MPS methods provide a tractable way to explore phase diagrams, compute ground states, and simulate real-time dynamics in spin chains and fermionic systems.

Long-range interactions, however, pose a well-known challenge for standard MPS implementations. A naive approach to incorporating extended couplings often requires rapidly growing bond dimensions, ultimately thwarting efficient simulations. Even the straightforward task of building the Hamiltonian in a suitable MPS-friendly format can become prohibitively expensive when interactions extend across large distances.

To address this issue in our fractional (long-range) Ising model, we employ a matrix product operator (MPO) representation of the Hamiltonian. Specifically, we approximate the fractional coupling profile by a finite sum of exponentials~\cite{wall2009emergent,wall2015out}, allowing the resulting MPO to remain at a manageable bond dimension. This construction preserves the computational benefits of MPS—chiefly their efficiency for states with limited entanglement—while enabling handling of genuinely long-range interactions. Consequently, we can accurately study the exotic nonlocal physics arising from fractional derivatives without the exponential overhead typical of unstructured long-range couplings.
In general, an MPO for a length-$L$ spin chain can be written as
\begin{equation}
    \label{eq:MPOdefinition}
    \hat{H}
    =
    \sum_{s_1,\ldots,s_L}
    \sum_{s_1^\prime,\ldots,s_L^\prime}
        W^{[1]}_{s_1,s_1^\prime}
        W^{[2]}_{s_2,s_2^\prime}
        \cdots
        W^{[L]}_{s_L,s_L^\prime}
    \,
    \lvert s_1,\ldots,s_L\rangle\langle s_1^\prime,\ldots,s_L^\prime \rvert,
\end{equation}
where each $W^{[j]}_{s_j,s_j^\prime}$ is a matrix of dimension $\chi_{j-1}\times\chi_j$ (the so-called ``virtual" bond dimension), and $s_j,s_j^\prime$ label the physical basis states (e.g.\ spin up/down). By maintaining relatively small bond dimensions $\chi_j$, an MPO provides a compact factorization of a potentially complicated and nonlocal many-body operator.

In our case, the spin--spin coupling $J(r)$ extends over long distances. A direct summation of all pairs $(i,j)$ in an operator form would require bond dimensions growing with system size. Instead, we exploit an exponential decomposition of form
\begin{equation}
\label{eq:ExponentSum}
    J(r)
    \approx
    \sum_{\alpha=1}^{N_{\mathrm{exp}}}
    a_\alpha\, e^{-b_\alpha\,r},
\end{equation}
where $r = |j-i|$, and the coefficients $\{a_\alpha, b_\alpha\}$ are chosen to fit $J(r)$ over the system size to a specified tolerance (e.g.\ $10^{-9}$). Each exponential term $e^{-b_\alpha r}$ admits a straightforward MPO representation with small bond dimension, and the total operator becomes an additive combination of these individually manageable pieces. Consequently, the bond dimension of the final MPO scales (at most) linearly in $N_{\mathrm{exp}}$ rather than in $L$, allowing for efficient tensor contractions in MPS-based time evolution or ground-state searches. In practice, approximately 10-14 terms are required in order to reach our tolerance to fit to $J(r)$.

This decomposition thus renders the fractional multiscale Hamiltonian amenable to established MPS/MPO techniques \cite{crosswhite2008applying}. Even though the physical interactions are long-range, the use of a finite sum of exponentials in Eq.~\eqref{eq:ExponentSum} yields an MPO of moderate bond dimension, thereby enabling simulations of larger systems at high accuracy without an exponential increase in computational cost.

Although the exponential sum in Eq.~\eqref{eq:ExponentSum} provides a compact operator representation, identifying the specific coefficients $\{a_\alpha,b_\alpha\}$ requires a robust fitting procedure. We adopt an iterative least-squares approach starting with a minimal two-term fit, then progressively adding one more exponential at each step until the maximum deviation from the exact fractional coupling $J(r)$ is below a target threshold (e.g.\ $10^{-9}$). At each iteration, the previously optimized parameters serve as the initial guess, which accelerates convergence and ensures a stable search in the high-dimensional parameter space.
\begin{figure}[t]
    \centering
    \includegraphics[width=\textwidth]{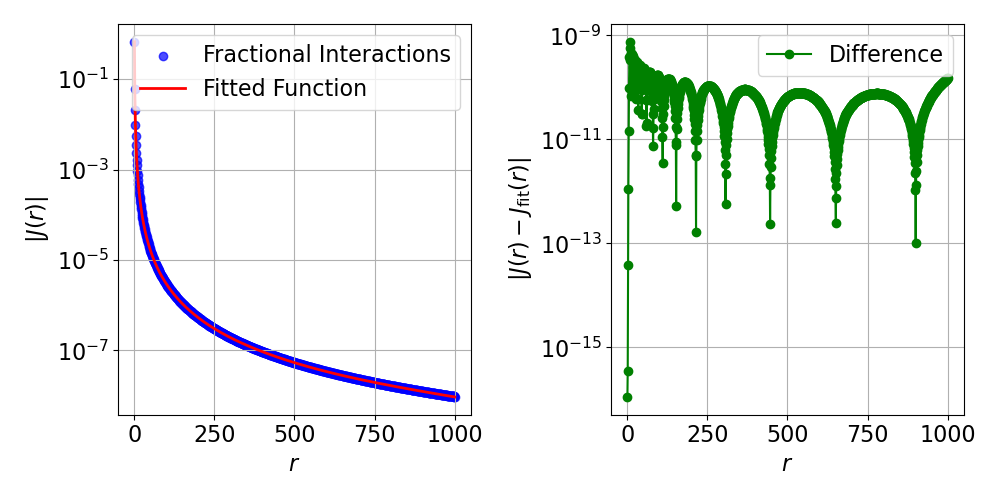}
    \caption{%
        Example of the iterative exponential decomposition for a fractional order $q=1.5$ on a chain of 1000 sites with a final finite sum of 12 decaying exponentials.
        (a)~Comparison of the original coupling $J(r)$ (circles) and its fitted sum of exponentials (solid line).
        (b)~Pointwise error $\Delta(r)$ between the exact and approximated profiles, showing that the error remains safely below the chosen threshold for all site separations $r$.
    }
    \label{fig:ExponentialDecomposition}
\end{figure}
~\Fref{fig:ExponentialDecomposition} illustrates the result of this procedure for $q=1.5$ and a system of 1000 sites. With each additional exponential term, both short-range features and long-distance tails are captured more accurately, making it possible to achieve high precision across the entire range of $r$.

\subsection{Ground State and Critical Point Computation}

To obtain the ground state of the fractional multiscale Ising Hamiltonian, we employ the \textsc{OpenMPS} library, which provides a highly optimized (Fortran-based) environment for constructing the ground state of a MPO defined Hamiltonian \cite{jaschke2018open, wall2015out}. This approach leverages translational invariance and adaptively increases the bond dimension as needed to capture the essential correlations of the system. 

Convergence is monitored through the energy variance, 
\begin{equation}
    \sigma_{E}^{2} = \langle \hat{H}^2\rangle - \langle \hat{H}\rangle^2 \,.
\end{equation}
The ground state search continues iteratively until $\sigma^2$ falls below a target threshold (here, $10^{-10}$). This variance makes a convenient measure of convergence as it directly determines an upper bound on the overlap between the numerically constructed ground state and states orthogonal to the true ground state \cite{jaschke2018open}.

Locating the quantum critical point (QCP) in the fractional multiscale Ising model requires scanning across different values of the transverse field $g$ and examining an order parameter that signals criticality. One might at first attempt to compute the energy gap directly. However this requires the high precision computation of not only the ground state, but also excited states. An often more practical criterion in one-dimensional systems is the bipartite entanglement entropy. In particular, at or near a QCP, the entanglement entropy typically reaches a pronounced maximum, reflecting strongly correlated degrees of freedom~\cite{wall2015out}. 

\begin{figure}[t]
    \centering
    \includegraphics[width=\textwidth]{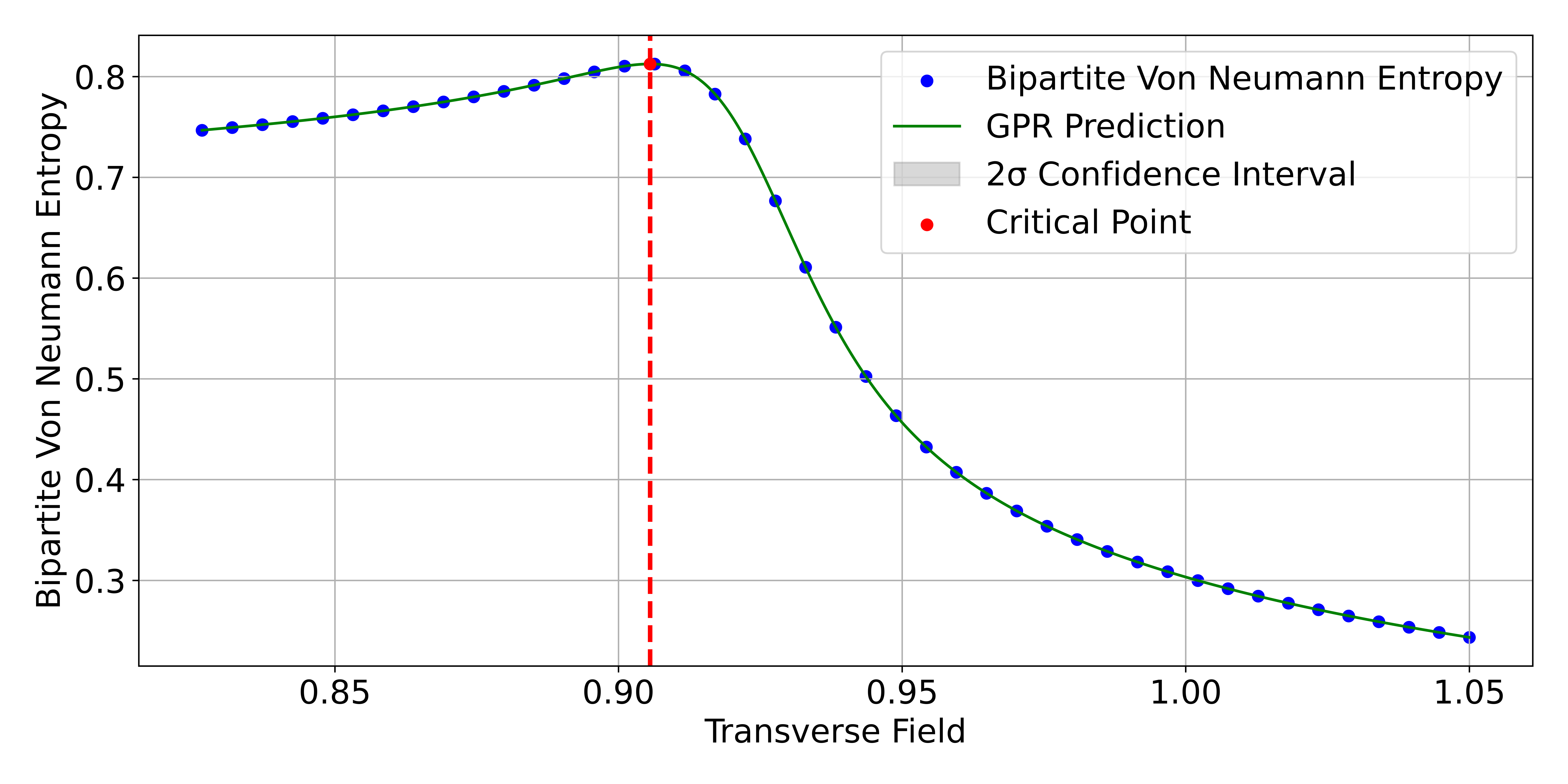}
    
    \caption{Bipartite entanglement entropy $S_{L/2}$ versus transverse field $g$ for a chain of $L=400$ sites at fractional order $q=1.5$. A clear maximum signals the critical point $g_c$. The solid line is a Gaussian process regression fit used to pinpoint the entropy peak from the discrete data points, providing a refined numerical estimate of $g_c$. Uncertainty associated with individual measurements was produced via repeated measurement and have a magnitude significantly smaller than displayed markers.}
    
    \label{fig:EntanglementEntropyExample}
\end{figure}

We thus prepare the ground state for a range of transverse field $g$ values, each time computing the bipartite entanglement entropy $S_{L/2}$ (the bond entropy), where the system is cut into two halves of equal size. ~\Fref{fig:EntanglementEntropyExample} shows an example of this procedure for $q=1.5$ and a chain of 400 sites. The entanglement entropy displays a clear peak at $g \approx g_c$, providing a numerical estimate for the critical transverse field. Once $g_c$ is identified, all subsequent dynamical studies focus on this parameter regime, where the system is most susceptible to long-range fluctuations and critical scaling. 
\subsection{Finite-Size Gap Analysis}

The dynamical critical exponent $z$ establishes how time scales diverge relative to spatial correlation lengths near a phase transition. If $\xi_{\text{space}}$ denotes the correlation length, then the corresponding correlation time $\xi_{\text{time}}$ scales as
\begin{equation}
    \xi_{\text{time}} \sim \xi_{\text{space}}^{\,z} \,.
    \label{eq:Definition_z}
\end{equation}
In a quantum many-body system, this relationship dictates the low-lying excitation spectrum. Near a critical point, $\xi_{\text{space}}\to\infty$ implies $\xi_{\text{time}}\to\infty$, leading to long-lived fluctuations.

One standard strategy to determine $z$ is to examine how the energy gap $\Delta$ scales with the system size $L$. In a finite system, let $\Delta(L)$ denote the gap between the ground state and the first excited state at or near the critical point. For large $L$, one typically expects a power-law decay,
\begin{equation}
    \Delta(L) \sim L^{-z} \,,
    \label{eq:gap_scaling}
\end{equation}
but subleading corrections can obscure the precise exponent if left unaccounted for. To mitigate this issue, we include a next-order term in the fitting function, for instance:
\begin{equation}
    \Delta(L) = a\,L^{-z}\left[1 + b\,L^{-\omega}\right] \,,
    \label{eq:gap_subleading}
\end{equation}
where $a,\,b,$ and $\omega>0$ are additional parameters determined by a global fit. This form helps separate universal scaling from non-universal finite-size effects, allowing for a more robust estimate of $z$.

A similar approach is used to extract the pseudocritical transverse field $g_c(L)$, which drifts toward the thermodynamic value $g_c$ as $L$ grows. Concretely, we fit
\begin{equation}
    g_c(L) = g_c + a\,L^{-1/\nu}[1 + b\,L^{-\omega'}],
\end{equation}
where $g_c$ is the asymptotic (infinite-size) critical point, and $\nu,\,a,\,b,\,\omega'$ encapsulate the leading and subleading scaling behavior. By applying these refined fitting forms to system sizes ranging from $L=10$ up to $L=200$, we are able to disentangle the universal features of the fractional multiscale Ising transition from spurious finite-size corrections, ultimately yielding more accurate values of both $z$ and $g_c$.

\begin{figure}[t]
    \centering
    \includegraphics[width=\textwidth]{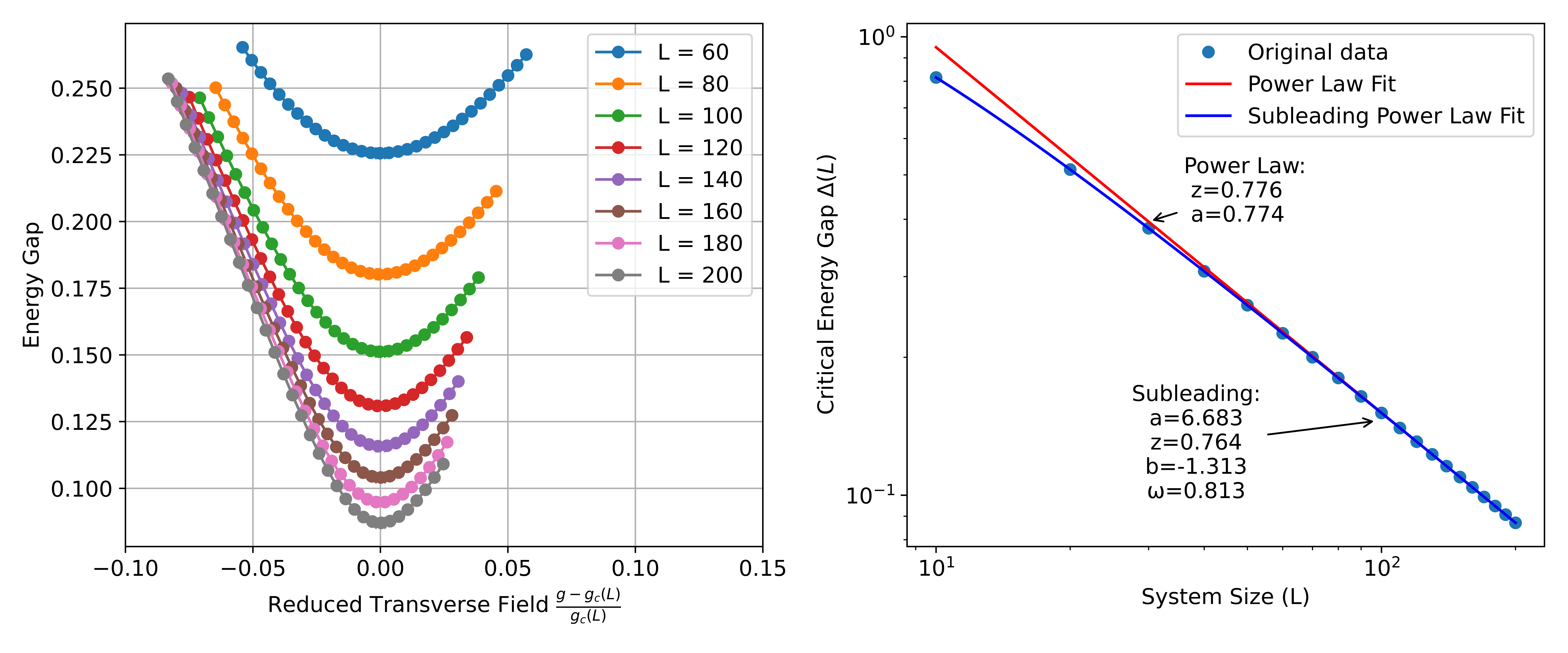}
    \caption{ Example fitting routine shown for a fractional order of 1.5 and linear system sizes from 10 to 200. (a) Energy gap $\Delta(L)$ versus transverse field $g$ for various chain lengths $L$. (b) Extracted dynamical exponent $z$ from a global fit of $\Delta(L)$ across multiple system sizes. Data points show numerical results, while the solid lines are fits to the leading power law and its subleading correction, as described in ~\eref{eq:gap_scaling} and \eref{eq:gap_subleading}. 
    }
    \label{fig:EnergyGapFitPlot}
\end{figure}

\subsection{Local Perturbation Time Evolution Analysis}

While finite-size gap analysis provides one measure of $z$, we also examine critical dynamics via a local perturbation protocol. Specifically, we initialize the system at the critical point $\left(g = g_c\right)$ in its ground state for a chain of $L = 200$ sites, then apply a time-dependent transverse bias along $\hat{\sigma}^z$ at the center of the lattice, at site $j^*$. From a quasiparticle perspective, this local injection of energy at criticality launches low-energy excitations whose velocity is governed by the dynamical exponent $z$. 

We modulate the perturbation amplitude $\lambda_0$ with a Blackman--Harris window $\omega_{\mathrm{BH}}(t)$. Over an interval $0 \le t \le \tau$, the perturbation Hamiltonian is
\begin{equation}
    \Delta \hat{H}(t)
    =
    \lambda_0 \,\omega_{\mathrm{BH}}(t)\,\hat{\sigma}_{j^*}^z \,.
\end{equation}
 Outside the time window $[0,\tau]$, we set $\omega_{\mathrm{BH}}(t) = 0$. Concretely, the Blackman--Harris function is
\begin{equation}
    \omega_{\mathrm{BH}}(t)
    =
    \begin{cases}
    a_0
    - a_1 \cos\left(\frac{2\pi t}{\tau}\right)
    + a_2 \cos\left(\frac{4\pi t}{\tau}\right)
    - a_3 \cos\left(\frac{6\pi t}{\tau}\right),
    & 0 \le t \le \tau,\\[6pt]
    0, & \text{otherwise} \,,
    \end{cases}
\end{equation}
with standard coefficients $\{a_0,a_1,a_2,a_3\}$ (e.g.\ $a_0=0.35875$, $a_1=0.48829$, $a_2=0.14128$, $a_3=0.01168$). This particular choice of coefficients strongly suppresses large-momentum (high-frequency) components, thereby allowing the perturbation to selectively excite the long-wavelength sector most relevant for uncovering universal scaling behavior \cite{harris2005use}.

Although one might track the propagation of a local perturbation by examining correlation functions or local magnetization, these observables often produce oscillatory interference patterns that obscure the light-cone boundary. In contrast, the bond entropy, $S_{\mathrm{bond}}(j,t)$, provides a comparatively monotonic signal of quasiparticle passage. As each quasiparticle moves through a bond, it entangles the two partitioned halves of the chain, causing $S_{\mathrm{bond}}(j,t)$ to rise above its baseline. Once the disturbance passes, the entropy saturates, producing a plateau within the light cone and leaving the outer region at its original, low-entropy value.

\begin{figure}[h]
    \centering
    \includegraphics[width=0.9\textwidth]{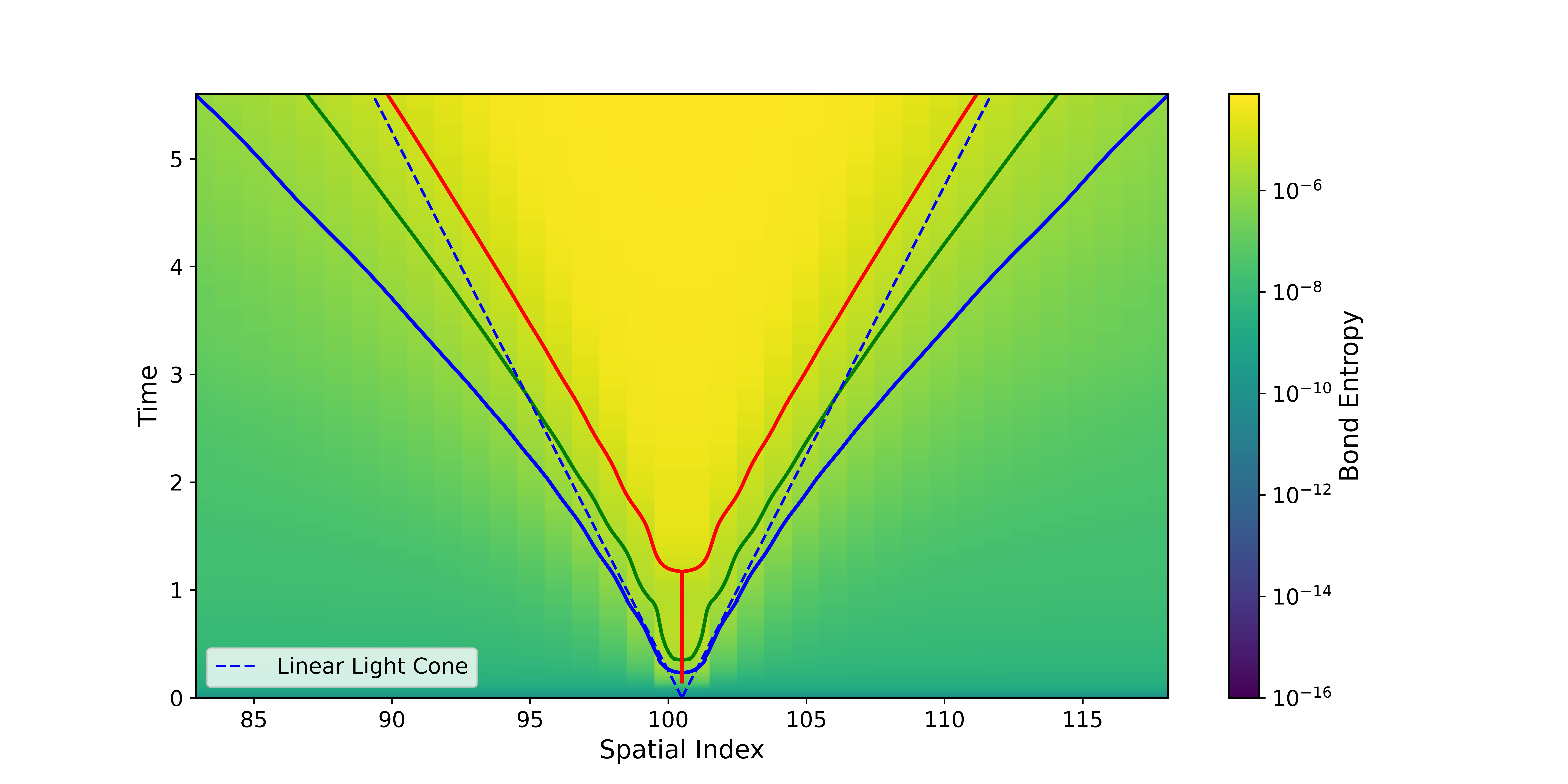}
    \\
    \includegraphics[width=0.9\textwidth]{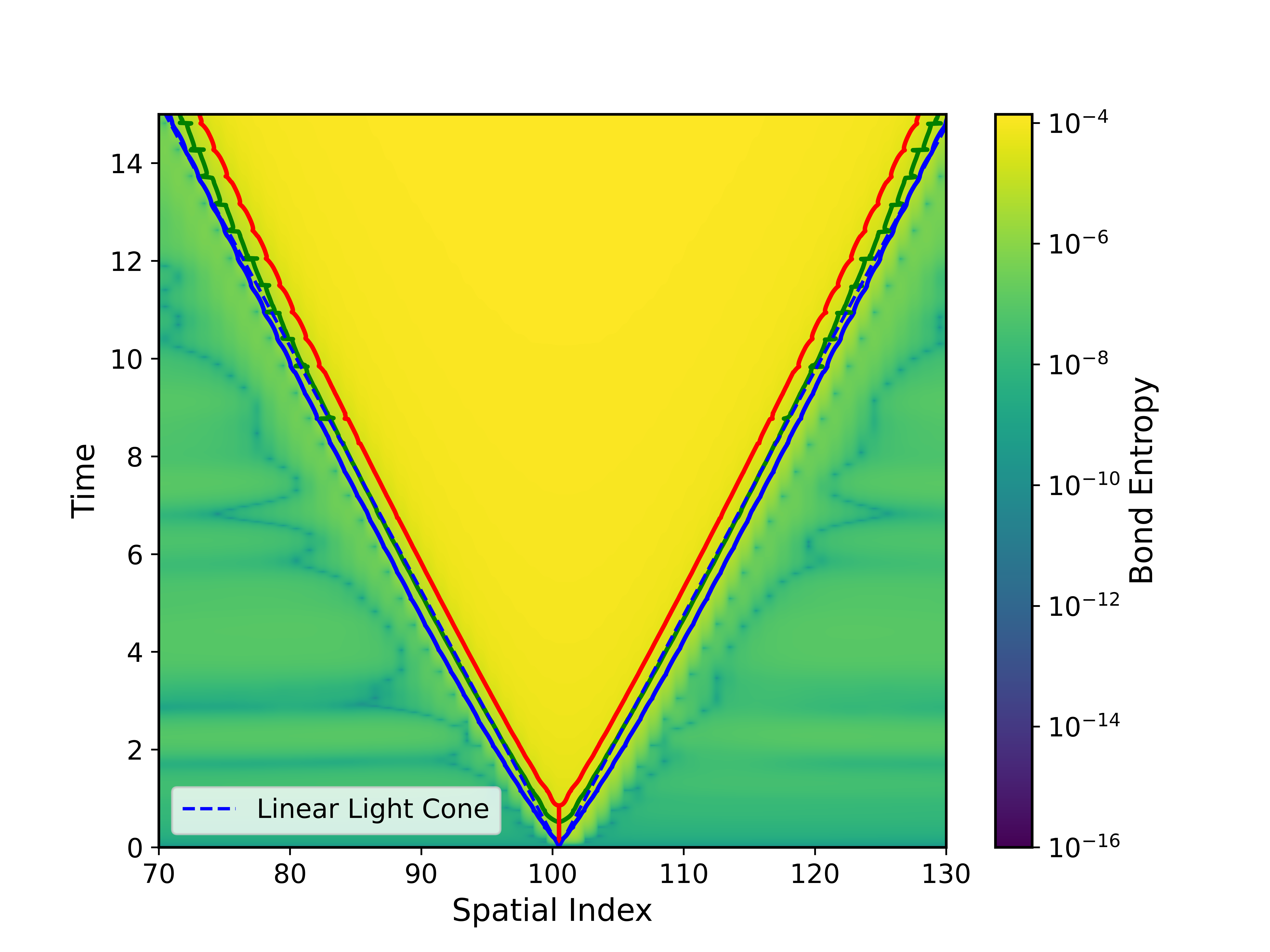}

    \caption{
        Bond-entropy contours $S_{\mathrm{bond}}(j,t)$ revealing the light-cone structure for fractional orders (a)~$q=1.5$ and (b)~$q=2.2$. In each panel, color indicates $S_{\mathrm{bond}}(j,t)$ as a function of position $j$ (horizontal axis) and time $t$ (vertical axis), measured after applying a local at site $j^*$. The solid contour lines correspond to three representative entropy levels, the propagation front of the disturbance. The dashed lines show a linear reference to emphasize how, for $q=1.5$, the light cone scales sublinearly, while for $q=2.2$, the cone is subtly superlinear and narrower than the linear reference.
    }
    \label{fig:BondEntropyContour}
\end{figure}

This behavior is particularly relevant in the fractional multiscale Ising model, where the usual short-range quasiparticle motion is replaced by Lévy-flight dynamics arising from the Riesz fractional derivative. By modifying the dispersion and speeds of low-energy excitations, such fractional effects reshape how entanglement spreads. Plotting $S_{\mathrm{bond}}(j,t)$ in the $(j,t)$-plane and identifying contours of constant entropy thus yields a direct visualization of the spreading of the initially local disturbance. Fitting these contours to
\begin{equation}
    |\,j - j^*\,|
    \sim
    t^{\frac{1}{z}}
\end{equation}
provides another robust estimate of the dynamical critical exponent $z$. Larger $z$ values correspond to slower-moving quasiparticles and narrower cones, while smaller $z$ values signify faster propagation and wider cones. An example for a fractional order $q=1.5$ and $q=2.2$ is shown in Figure.~\ref{fig:BondEntropyContour}.

\section{Truncated Jordan-Wigner Mapping: Mean Field Diagonalization}

To gain further insight into how low-energy excitations shape the dynamical critical exponent, we transform the fractional Ising model via the Jordan–Wigner transformation, followed by Fourier and Bogoliubov transformations. In a mean-field approach, fractional interactions manifest as momentum-dependent dispersion terms, clarifying how nonlocal couplings modify the quasiparticle spectrum. 

The Jordan--Wigner mapping identifies each spin operator with fermionic creation and annihilation operators $\hat{c}_i^\dagger, \hat{c}_i$, satisfying

\begin{equation}
\{\hat{c}_i,\,\hat{c}_j^\dagger\} = \delta_{ij}, \quad \{\hat{c}_i,\,\hat{c}_j\} = 0 = \{\hat{c}_i^\dagger,\hat{c}_j^\dagger\} \,.
\end{equation}
In particular, one can write

\begin{equation}
\hat{\sigma}_j^z = 1 - 2 \hat{c}_j^\dagger \hat{c}_j, \quad \hat{\sigma}_j^x = (\hat{c}_j^\dagger + \hat{c}_j) \prod_{k < j} \hat{\sigma}_k^z \,,
\end{equation}
where the latter product (often expressed as a Jordan-Wigner string) ensures the correct anticommutation relations between neighboring spins. Substituting into~\eref{eq:FracIsing}, 

\begin{equation}
\hat{H}' = -J_0 \sum_{i<j} J(|j-i|) (\hat{c}_i^\dagger + \hat{c}_i) \left[\prod_{k=i+1}^{j-1} (\hat{\sigma}_k^z)\right] (\hat{c}_j^\dagger + \hat{c}_j) + g \sum_j (1 - 2\hat{c}_j^\dagger\hat{c}_j) \,.
\end{equation}
In the standard Jordan--Wigner transformation, the operators $\sigma_i^x$ carry a string of $\sigma_k^z$ factors for $k<i$. These strings ensure the correct fermionic anticommutation relations, but they also introduce higher-order (beyond quadratic) coupling terms when two spins at sites $i$ and $i+r$ interact. To simplify the fractional multiscale Ising Hamiltonian, one can truncate these non-quadratic contributions by effectively replacing the Jordan--Wigner strings with unity. Physically, this neglects nonlocal fluctuations not captured by a uniform magnetization. Because the bilinear terms already fix the softest dispersion, any $2n>2$ fermion vertex generated by the Jordan–Wigner string carries extra powers of the bilinear propagator and therefore a higher canonical dimension as long as we are within the mean field window for the model. As such, the result is consistent with a mean field dynamical critical exponent.  Under the renormalization group rescaling $x\to b\,x, \tau \to b^{z}\tau$, these higher‑order operators become irrelevant to the leading space–time anisotropy; they merely renormalize non‑universal amplitudes and do not alter the exponent determined from the quadratic term \cite{defenu2019dynamical}. Truncating $\hat{H}'$ to only bilinear terms simplifies to

\begin{equation}
    \hat{H}' \approx -J_0\sum_{j}\sum_{r>0} J(r) \left(\hat{c}_j^\dagger\hat{c}_{j+r}^\dagger + \hat{c}_j^\dagger\hat{c}_{j+r} + \hat{c}_{j+r}^\dagger\hat{c}_j + \hat{c}_{j+r}\hat{c}_j \right) + g\sum_{j} \left(1 - 2\hat{c}_j^\dagger\hat{c}_j\right) \,,
\end{equation}
consistent with the long-range Kitaev model solved exactly in \cite{jaschke2017critical, vodola2014kitaev}. A standard Fourier transform for the creation and annihilation operators is

\begin{equation}
    \hat{c}_j = \frac{1}{\sqrt{N}}\sum_{k} e^{i\,k\,j}\hat{c}_k, \quad \hat{c}_j^\dagger = \frac{1}{\sqrt{N}}\sum_{k} e^{-i\,k\,j}\,\hat{c}_k^\dagger \,,
\end{equation}
where $j$ labels sites, $k$ is the momentum, and $i=\sqrt{-1}$. Substituting these into our fermionic creation and annihilation operators yields a quadratic expression in momentum space. Each term, for example

\begin{equation}
\hat{c}_j^\dagger \hat{c}_{j+r} = \frac{1}{N}\sum_{k,q} e^{-i\,(k-q)\,j}e^{i\,q\,r}\,\hat{c}_{k}^\dagger\hat{c}_{q} \,,
\end{equation}
becomes part of a double sum over $j, r$. Crucially, summing over $j$ enforces 

\begin{equation}
\sum_{j} e^{-i\,(k-q)\,j} = N\delta_{k,q}
\end{equation}
which collapses the sums over $k$ and $q$ into a single momentum index. The resulting terms form a purely bilinear Hamiltonian in momentum space

Collecting all diagonal and off-diagonal terms in momentum space, we obtain

\begin{equation}
\hat{H}' \approx \sum_{k}[\xi_k\hat{c}_k^\dagger \hat{c}_k + \frac{1}{2}\Delta_k (\hat{c}_k^\dagger \hat{c}_{-k}^\dagger + \hat{c}_{-k} \hat{c}_{k})] + \text{const.} \,,
\end{equation}
where

\begin{equation}
    \xi_k= 2g - 2J_0 \sum_{r>0} J(r)\cos(k\,r), \quad \Delta_k = 2J_0 \sum_{r>0} J(r)\sin(k\,r) \,.
\end{equation}
One then performs a standard Bogoliubov transformation to diagonalize the Hamiltonian. Defining new quasiparticle operators $\hat{\gamma}_k,\hat{\gamma}_k^\dagger$ through a rotation in each momentum sector, the final result is

\begin{equation}
    \hat{H}' = \sum_{k} E_k \left(\hat{\gamma}_k^\dagger \hat{\gamma}_k - \frac{1}{2}\right),
\end{equation}
where the Bogoliubov dispersion is 

\begin{equation}
    E_k = \sqrt{\xi_k^2 + \Delta_k^2} \,.
\end{equation}
Labeling these trigonometric summations as,

\begin{equation}
    \mathcal{C}_k= J_0\sum_{r>0} J(r)\cos(k\,r), \quad \mathcal{S}_k = J_0\sum_{r>0} J(r)\sin(k\,r) \,,
\end{equation}
the dispersion term may be rewritten as

\begin{equation}
    E_k = 2\sqrt{\mathcal{C}_k^2 + \mathcal{S}_k^2 + 2g\mathcal{C}_k + g^2} \,.
    \label{eq:bogoliubovEnergy}
\end{equation}
Remarkably, the sum

\begin{equation}
    \sum_{r>0} J(r)\cos(k\,r)
\end{equation}
admits a closed-form expression once rewritten in terms of complex exponentials over all integer distances. Extending the summation to negative r and absorbing $r=0$ into a global shift, 

\begin{equation}
    \sum_{r>0} J(r)\cos(k\,r) = \frac12 \sum_{r=-\infty}^{\infty} J(|r|)e^{i\,k\,r}
    -\frac12\,J(0) \,.
\end{equation}
As shown by Ortigueira in \cite{ortigueira2006riesz},

\begin{equation}
    \sum_{r=-\infty}^\infty (-1)^r\binom{q}{q/2+r}e^{ikr} = \left|2\sin\left(\frac{k}{2}\right)\right|^q \,,
\end{equation}
which is precisely the fractional multiscale Ising interactions J(r) applied as a cosine summation, yielding

\begin{equation}
    \mathcal{C}_k= \frac{1}{2}\left(J_0 J(0) - \,J_0 \left|2\sin\left(\frac{k}{2}\right)\right|^q  \right)\,. 
\end{equation}
Although the cosine series admits a simple closed-form solution, the sine series involves the imaginary part of a hypergeometric function. Specifically, both sine and cosine summations can be compactly written in terms of

\begin{equation}
    \begin{aligned}
       & \frac{\Gamma(q+1)}{\Gamma(q/2+1)^2} \,{}_2F_{1}(1,\,-\,\frac{q}{2};\,\frac{q+2}{2};\,e^{\,i\,k}) = \\
        &=\sum_{r=0}^\infty  (-1)^r\binom{q}{q/2+r}\cos(kr) + i  (-1)^r\binom{q}{q/2+r}\sin(kr) = F(q,k)
    \end{aligned}
\end{equation}
where the real part reproduces the cosine sum and the imaginary part gives the sine sum. Unlike its cosine counterpart, however, the sine summation does not collapse to a simple power-law expression and retains a hypergeometric form

\begin{equation}
    \begin{aligned}
        \mathcal{C}_k/J_0&=-\Re\left[\frac{\Gamma(q+1)}{\Gamma(q/2+1)^2} \,{}_2F_{1}(1,\,-\,\frac{q}{2};\,\frac{q+2}{2};\,e^{\,i\,k})\right] +\binom{q}{q/2} \,, \\
        \mathcal{S}_k/J_0 &= \Im\left[\frac{\Gamma(q+1)}{\Gamma(q/2+1)^2} \,{}_2F_{1}(1,\,-\,\frac{q}{2};\,\frac{q+2}{2};\,e^{\,i\,k})\right] \,.
    \end{aligned}
    \label{eq:CkSkHypergeometricDefinition}
\end{equation}
Substitution of~\eref{eq:CkSkHypergeometricDefinition} into~\eref{eq:bogoliubovEnergy} yields
\begin{equation}
    E_k=\sqrt{\left|F(q,k)\right|^2 + \left[\binom{q}{q/2}-g\right]\left|\sin\left(\frac{k}{2}\right)\right|^q +g\left[g-\binom{q}{q/2}\right]} \,.
\end{equation}
We then observe a critical point associated with momenta k=0 for $g_c=\frac{1}{2}\binom{q}{q/2}$. To derive the dynamical critical exponents, we observe how energy scales with the lowest energy perturbations (the limit of $k\to0$ at $g_c$). 
\begin{equation}
    E_k=\sqrt{\left|F(q,k)\right|^2  - \left[\frac{1}{2}\binom{q}{q/2}\right]^2 + \left[\frac{1}{2}\binom{q}{q/2}\right]\left|\sin\left(\frac{k}{2}\right)\right|^q} \,.
\end{equation}
We then note that the hypergeometric function term approaches exactly $\frac{1}{2}\binom{q}{q/2}$ as k approaches 0 such that,
\begin{equation}
    \lim_{k\to0}\left|\frac{\Gamma(q+1)}{\Gamma(q/2+1)^2} \,{}_2F_{1}(1,\,-\,\frac{q}{2};\,\frac{q+2}{2};\,e^{\,i\,k})\right| - \left(\frac{1}{2}\binom{q}{q/2}\right) = 0 \,.
\end{equation}
By then performing a Euler transform we may obtain exactly the leading power law behavior of the hypergeometric function as we approach $k=0$ ($z=e^{ik}=1$) given by,
\begin{equation}
    {}_2F_1(a,b;c;z) = (1-z)^{c-a-b}{}_2F_1(c-a,c-b;c,z) \,.
\end{equation}
This transform extracts the leading scaling behavior and implies
\begin{equation}
    \left|\frac{\Gamma(q+1)}{\Gamma(q/2+1)^2} \,{}_2F_{1}(1,\,-\,\frac{q}{2};\,\frac{q+2}{2};\,e^{\,i\,k})\right|^2 - \left(\frac{1}{2}\binom{q}{q/2}\right)^2 \propto |k|^{2q} \,.
\end{equation}
However, it is the second term that applies the dominant scaling behavior of $E_k$ as,
\begin{equation}
    \frac{1}{2}\binom{q}{q/2}\left|\sin\left(\frac{k}{2}\right)\right|^q \propto|k|^q \,,
\end{equation}
implying that with small perturbations at exactly the critical point, the fractional multiscale Ising model scales with momenta as
\begin{equation}
    E_k\propto |k|^{q/2}\,,
\end{equation}
so that in the mean-field approximation, the model already predicts
\begin{equation}
    z=\frac{q}{2} \,.
    \label{eq:dynamicZResult}
\end{equation}

\section{Anomalous Dynamical Scaling and L\'evy Causality}

\begin{figure}[h]
    \centering
    \includegraphics[width=0.5\textwidth]{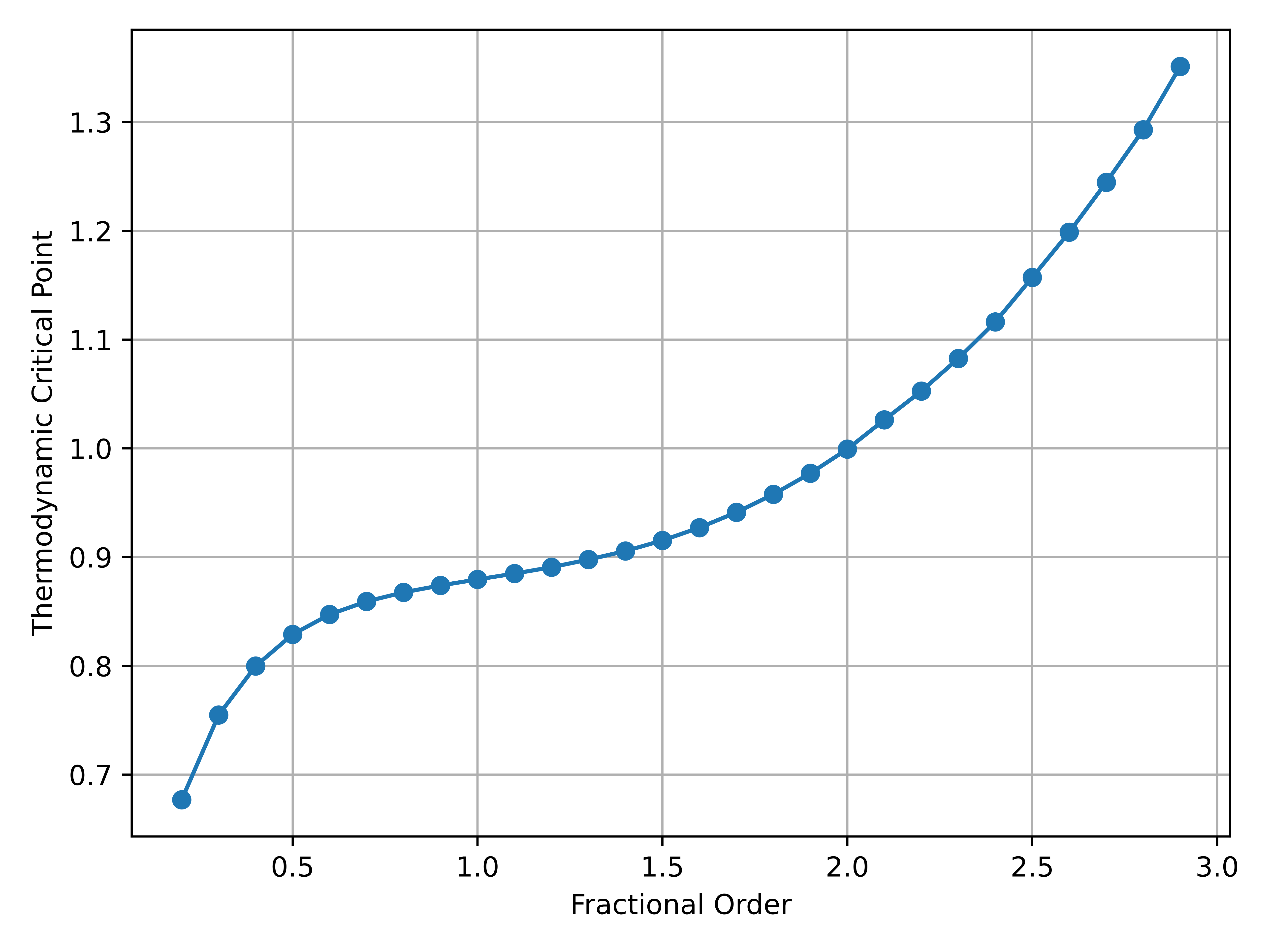}
    \caption{
    Thermodynamic critical transverse field $g_c$ versus fractional order $q$, determined by finite-size scaling of chains up to $L=200$. The data points show a monotonic increase in $g_c$ with growing $q$. Beyond $q>2$, where frustration emerges from sign-alternating long-range interactions, larger transverse fields are necessary to disrupt the ordered phase. Error bars are smaller than markers. 
    }
    \label{fig:CriticalPointVsOrder}
\end{figure}

\Fref{fig:CriticalPointVsOrder} traces the critical transverse field $g_c$ in the thermodynamic limit as a function of the fractional order $q$. The threshold rises monotonically: for $q<2$ the chain retains genuine power-law couplings, while at $q=2$ it reduces to the nearest-neighbour transverse-field Ising model and reproduces the textbook critical point. When $q>2$ the spectrum is governed by effectively local ferromagnetic bonds frustrated by a weak long-range antiferromagnetic tail, so the transition persists yet shifts to larger $g_c$. The reflection formula applied to the kernel in Eq.~\ref{eq:coupling} shows that the integrated coupling strength grows with $q$; even after enforcing a Kac normalization that fixes the total weight to unity, the field required to close the gap continues to increase as a result of long range interactions beyond quadratic after a Jordan-Wigner transform \cite{pfeuty1970one}. Intuitively, the emergent antiferromagnetic tail for $q>2$ counteracts the dominant ferromagnetic alignment, demanding a stronger transverse field to disorder the spins and thereby elevating $g_c$.

\begin{figure}[h]
    \centering
    \includegraphics[width=\textwidth]{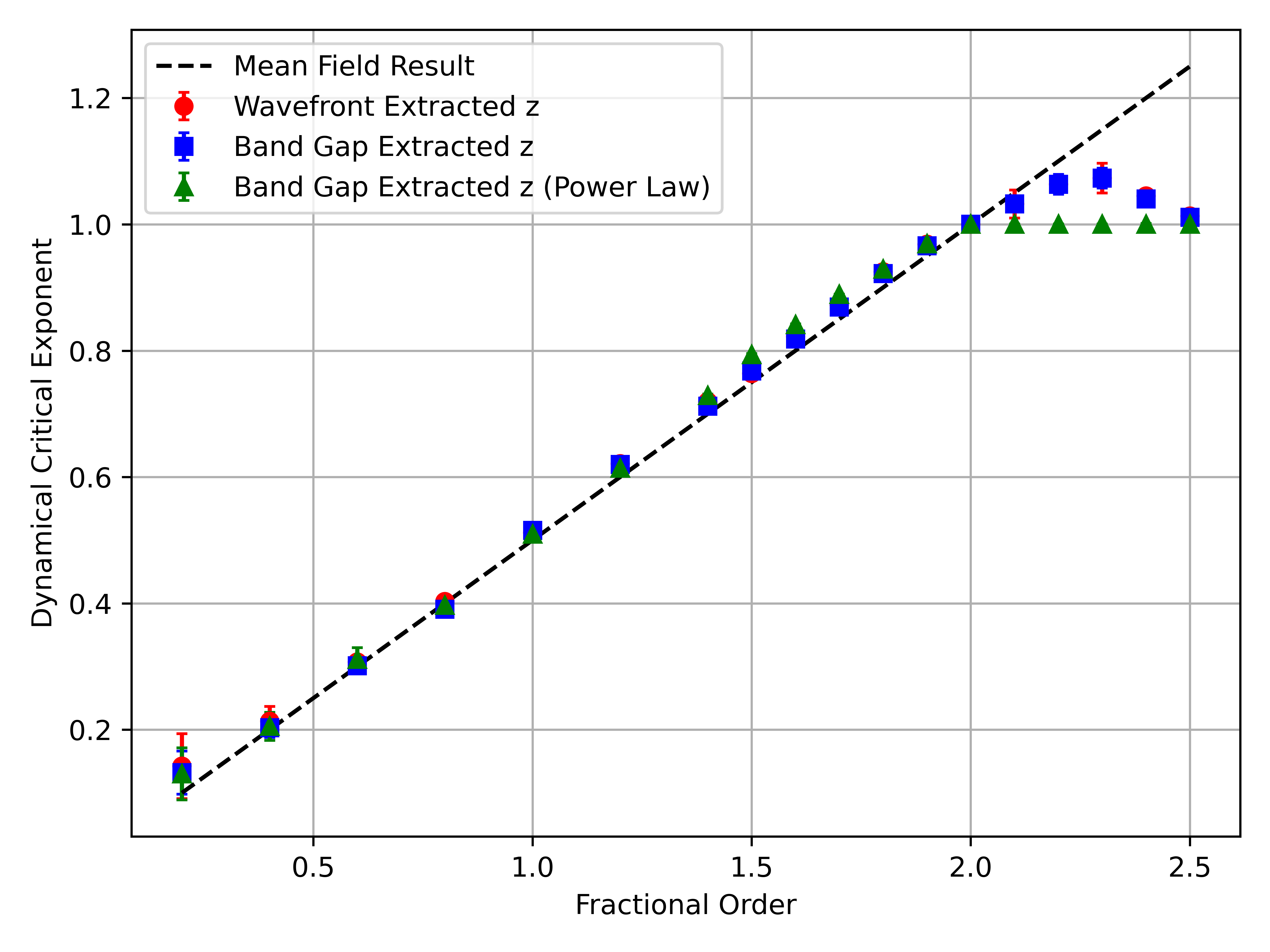}
    \caption{
        Dynamical critical exponent $z$ as a function of the fractional order $q$. Blue markers denote data from finite-size gap scaling, while red markers come from local-perturbation wavefront analysis (bond-entropy contours). The black dashed line indicates the mean field prediction $z = \frac{q}{2}$. Finally, green markers denote the calculated dynamical critical exponent for the asymptotic power law of the fractional interactions. For $q > 2$, $z$ initially exceeds 1, producing superlinear light cones arising from the frustration from sign-alternating interactions. However, at larger $q$, the antiferromagnetic couplings cease to affect the critical scaling, causing $z$ to drift back toward unity leading to the system to become effectively local again.
    }
    \label{fig:ZvsQ}
\end{figure}

We next examine the dynamical critical exponent $z$ using two complementary methods: (1) finite-size gap scaling and (2) local-perturbation wavefronts. ~\Fref{fig:ZvsQ} summarizes our main findings. For $0 \leq q \leq 1$, both methods agree within an uncertainty of 0.01, that $z \approx \frac{q}{2}$, closely matching earlier analytical predictions for power-law Ising chains~\cite{maghrebi2016causality}. This relationship implies that reducing $q$ lowers the effective dispersion relation of the model, speeding up quasiparticles, then at $q=2$, we recover the conventional nearest-neighbor universality.

However, as the fractional order increases, the extracted dynamical critical exponents observe a distinct deviation from the mean-field prediction as fluctuations become significant, and faithful finite-size scaling demands much larger lattices to uncover the true asymptotic behaviour. As an additional check, we extracted z from the asymptotic power-law dependence of the band gap. This independent estimate agrees with the wavefront analysis to within 5\% across all q, with the largest discrepancies appearing in the fluctuation-dominated regime.

For $q>2$, we still find a well-defined scaling form for $z$, but the system no longer maps smoothly onto standard power-law models. Instead, the frustrated interactions induce a superlinear light-cone that lies outside of the universality class associated with simple power law models. Importantly, at larger fractional orders (beyond $q \gtrsim 2$), our results also reveal a distortion in the dynamical critical exponent possibly due to crossover effects toward an effectively local regime, especially in one dimension. Although $z$ remains close to $q/2$ in principle, the low-dimensional setting amplifies corrections from finite-size and frustration-driven crossovers, causing a nontrivial drift in the exponent.

A useful point of comparison arises in the standard transverse-field Ising model, where introducing a transverse field at criticality effectively upgrades the low-energy description to a relativistic-like conformal field theory with $z=1$. In the absence of a transverse field, pure Ising interactions simply disperse excitations with mean-square displacement proportional to $t$, indicative of first-order time evolution. The addition of a critical transverse field modifies the effective equation of motion, furnishing a second-order time derivative and driving the system to a universality class characterized by $z=1$. 

In the fractional multiscale transverse‑field Ising model, a mean field analysis yields a dynamical exponent $z = q/2$ as shown in~\eref{eq:dynamicZResult}. Consequently, whereas a wave packet governed by the fractional Schrödinger equation spreads sub‑ballistically as $r(t) \propto t^{1/q}$, the emergent Lorentz‑like symmetry at the Ising critical point doubles this exponent, producing a elevated scaling with $r(t) \propto t^{2/q}$ exactly as with the standard $q=2$ case. Such behavior is consistent with how fractional derivatives naturally connect to Lévy-flight statistics, providing a more fundamental foundation for analyzing nonlocal quantum criticality. 

Our results can be compared with the two‑loop renormalization‑group study of long‑range criticality in Ref.~\cite{maghrebi2016causality}. That work demonstrated that, for couplings $J(r)\sim r^{-(1+q)}$ in one dimension, the mean‑field relation $z = q/2$ holds only within the strictly long‑range window $0 < q < 2/3$; for $q > 2/3$ critical fluctuations introduce nonlinear corrections that bend $z$ away from $q/2$ and drive the flow toward short‑range behaviour with $z \to 1$. We hypothesise that these discrepancies arise from finite‑size effects, and that at larger system sizes fluctuation‑driven corrections will diminish the apparent mean‑field window, bringing the crossover closer to the two‑loop prediction.

Although a superlinear light-cone is evident for $q>2$, it remains relatively subtle in one dimension. Because the sign-alternating couplings still decay as $1/r^{1+q}$, their frustration effect is significant enough to distort the universality class yet not overwhelming; the resultant superlinear cone ultimately competes with the short-range ferromagnetic response. In higher-dimensional fractional multiscale models, however, we expect this frustrated regime to become more pronounced, as the number of such long-range, sign-changing bonds grows faster with system size. This could enhance the superlinear behavior and potentially bring the numerical results into closer alignment with mean-field predictions. Exploring higher-dimensional fractional multiscale Ising systems is thus a natural next step to fully uncover the interplay between frustration, heavy-tailed couplings, and anomalous dynamical scaling.

\section{Conclusion}
Fractional derivatives establish a connection between heavy‑tailed L\'evy‑flight statistics and quantum criticality, extending single‑particle formulations~\cite{laskin2000fractional,laskin2002fractional} to many‑body spin systems.  For $q<2$, our simulations confirm that the effective quasiparticle dispersion reconstructs the dynamical exponent within 5\% of $z=q/2$, matching known results for power‑law interactions, whereas for $q>2$ sign‑alternating couplings emerge and drive a distinct, super‑linear light cone.  This continuous crossover from near power‑law physics ($q<2$) to a non‑local, frustrated regime ($q>2$) provides a theoretical richness of the fractional multiscale Ising model and shows how tuning the fractional order $q$ unlocks unconventional phase structures beyond both nearest‑neighbour and conventional long‑range systems. 

A natural next step is to lift the model to higher dimensions.  In $d\ge2$ the larger phase‑space volume suppresses the one‑dimensional infrared fluctuations that cause deviations from mean‑field theory; consequently we expect the mean‑field long-wavelength prediction $z=q/2$ to observe less deviation throughout the $q>2$ regime, producing an even more pronounced superlinear "Lévy light cone" that could help confine information propagation and, by reducing available transport channels, mitigate decoherence in planar or three‑dimensional qubit arrays.  Fractional extensions in 2D and 3D therefore offer a controlled arena for studying unique quantum dynamics and may guide the design of novel architectures that exploit frustrated, non‑local couplings.

On the experimental front, modern quantum simulators—such as trapped‑ion chains~\cite{Richerme2014}, programmable Rydberg‑atom arrays~\cite{Bernien2017}, and circuit‑ or cavity‑QED architectures with photon‑mediated couplings~\cite{Sundaresan2019}—already enable approximate realizations of fractional kernels. These platforms could directly probe the predicted Lévy‑like light cones and anomalous critical exponents. Taken together, the fractional multiscale Ising framework unifies short‑range, extended‑range and frustrated interactions in a single tunable model, provides fresh insight into nonlocal models in terms of Lévy statistics, and opens multiple experimental and theoretical avenues for exploring non‑local quantum matter.

\bibliography{sources}

\end{document}